\documentclass[notoc]{JHEP} 
\usepackage{amsmath}
\usepackage{epsfig}
\usepackage{amssymb,amsfonts}

\def\be{\begin{equation}}
\def\ee{\end{equation}}
\def\ba{\begin{eqnarray}}
\def\ea{\end{eqnarray}}

\def\tr{\,{\rm tr}\, }

\newcommand{\bR}{{\bf R}}

\def\sqr#1#2{{
\vcenter{\vbox{\hrule height.#2pt
\hbox{\vrule width.#2pt height#1pt \kern#1pt
\vrule width.#2pt}
\hrule height.#2pt}}}}
\boldmath
\title{D-branes in some  near-horizon geometries}
\unboldmath
\author{C. Bachas\\
Laboratoire de Physique Th{\'e}orique
de l'Ecole Normale Sup{\'e}rieure\thanks{Unit{\'e} mixte  du
CNRS et de l'Ecole Normale Sup{\'e}rieure,  
UMR 8549.} \\ 
24  rue Lhomond, 75231 Paris Cedex 05, France\\
\email{bachas@physique.ens.fr}}

\abstract{I review some  properties of D-branes in the SU(2) and
SL(2,R) WZW models. I comment on a potential difficulty for the
realization of `warped brane worlds' in string theory. This short note is based
on a  talk given  at  the Strings'01 conference in Mumbay. 
} 

\begin{document}
\section{Introduction}

In this talk I will
review some properties of D-branes in the
SU(2) and SL(2,R) WZW models. These models enter in  some of
the earliest,  {\it exact} and stable string-theory
backgrounds \cite{or,abs,chs,hor}, 
that   have returned to center-stage recently  in the light of
the AdS/CFT correspondence \cite{ma,revN}. 
Specifically, the level-$k$ SU(2) model describes,  together with a Feigin-Fuchs
field, the near-horizon geometry of $Q_5 = k+2$ Neveu-Schwarz fivebranes \cite{chs}.
The background  near the horizon of $Q_5$ NS5-branes and $Q_1$ fundamental
strings involves, on the other hand,  both an SU(2) and an  SL(2,R) WZW model 
(see for example \cite{ms}).
In these settings one hopes, in principle, to go beyond the gauge theory/gravity
correspondence, and  test the conjectured duality between
a full string theory and a field theory.

  The study of D-branes in these backgrounds is important
for a variety of reasons: (a) D-branes are an essential
ingredient of the corresponding string theories, 
and have  interesting holographically
dual interpretations; (b) despite substancial
 progress,\footnote{In reference \cite{mo}, in particular, a complete proposal was
made for the spectrum of closed-string excitations. A recent review
and many more references on the SL(2,R)  model is  \cite{mar}.}
the SL(2,R) WZW model is still only partially understood.
The semiclassical analysis of D-branes could help `solve'
this important CFT; and (c) these backgrounds are controllable
playgrounds,  in which to see whether  `warped brane world' scenarios \cite{RS}
can be  realized in string theory.   

 I will barely  touch upon these questions  in my talk.
For lack of space, I will not even 
mention many  works that have analyzed 
branes in near-horizon
geometries with Ramond-Ramond fluxes. 
My  `excuse' is that these do not have, at present,  
an exact CFT description. Finally, as this talk was being written,
there have appeared several papers discussing  related  issues
\cite{MMS,KR2,KR1,PR,GKS,Hi}. I will comment on them  succintly in the
appropriate places.


\boldmath
\section{D-branes in SU(2)}
\unboldmath

 According to Cardy's general prescription \cite{Cardy}, 
applicable to any rational CFT, the basic conformal boundary states 
of the SU(2) WZW model are
\begin{equation}
\vert\; n \gg_C \;  =\;  \sum_{m=1}^{k+1}\; 
\frac{S_n^{\ m}}{\sqrt{S_1^{\ m}}}\;  \vert\; m \gg_I \ .
\end{equation} 
Here $k$ is the level of the current  algebra, and 
$S_n^{\ m}$ is the modular-transformation matrix for the 
characters $\chi_m$. The representations of the 
chiral algebra are labelled
by the dimension of the highest-weight
subspace  $m=1,\cdots ,  k+1$ .  
The `character' 
or Ishibashi states  $\vert m \gg_I$  only couple
to closed-strings 
 in the representation ${\cal H}_m \otimes {\bar {\cal H}}_m $,
and with unit stength. Using Verlinde's formula one can transform  the
cylinder diagram, describing the exchange of closed strings
between the $n$ and $n^\prime$ Cardy states, to  
the open-string (annulus)   channel with the result
\begin{equation}
{\cal A}_{n n^\prime}\;  =\;
 \sum_{r}  {\cal N}_{n n^\prime}^{\ r}\; \chi_r(q)\ , 
\end{equation} 
where  ${\cal N}_{n n^\prime}^{\ r}$ are the 
fusion coefficients. These are
non-negative integers, as required by the identification
of Cardy states with regular  D-branes.

Although this algebraic construction was  known for several  years, 
its geometric meaning has been  only  clarified recently.
A  simple but  illuminating remark  is that the identification
of left and right currents on the worldsheet boundary, which is
automatically imposed by all Cardy  states,   
translates  into  Dirichlet conditions for directions normal to
conjugacy classes of the group \cite{AS1}. 
Explicitly, if we parametrize the strip worldsheet by $(\sigma,\tau)$,
then the identification $J^a = {\bar J^a}$  implies  
\begin{equation}  
    \left[ 1+ {\rm Ad}(g) \right]\; g^{-1} \partial_\tau g\;  =\; 
  \left[ 1 - {\rm Ad}(g) \right]\;  
g^{-1} \partial_\sigma g\ .
\label{gl}
\end{equation}
Here ${\rm Ad}(g) L \equiv 
 g L g^{-1}$, so that $\left[ 1 - {\rm Ad}(g) \right]$
projects  onto the tangent space of the  conjugacy class of $g$.
It follows that the worldsheet boundaries, parametrized by 
the coordinate $\tau$,
are stuck on conjugacy classes of the group. 
In the case of SU(2), these  
are spherical D2-branes.

  What stops these branes from shrinking to a point is a
quantized worldvolume flux $\int F$, whose  interaction  with the
background Neveu-Schwarz $B$-field is described by  the
 Dirac-Born-Infeld (DBI) action (see for instance \cite{revs}).
The  semiclassical
analysis based on this  action \cite{BDS} 
(see also \cite{Paw}) reproduces many  detailed properties
of the SU(2) D-branes,  and offers a nice geometric interpretation of
the algebraic data of this  CFT. 
Furthermore, most  of these  semiclassical
results turn out to be  {\it exact}, to all orders in the
$\alpha^\prime$ expansion,   thanks presumably
 to the existence of supersymmetric
embeddings. I will not review these calculations here -- the reader
can consult the original articles.

   Let me instead focus on the  subtle question   of how to
define the  Ramond-Ramond charge of  these D-branes. The issue 
has been elucidated  from various angles in references 
\cite{K1,Eli,K2,K3,K4,K5,K6}. The two natural candidates \cite{BDS}
for the charge are:  
(i) the integral of the two-form
${\cal F} =  \hat B + 2\pi\alpha^\prime F$, which is gauge-invariant
but not quantized; and (ii) the integral of $F\simeq dA$ which
is quantized, but changes under large gauge transformations
$\delta B\simeq d\Lambda$. 
As it turns out, 
the former gives  the local ('source') coupling to  RR  fields,
 while the latter, after periodic identification modulo $k+2$, is
the invariant charge of  a 
twisted version \cite{BM,K6} of K theory 
\cite{MM,W1,Wrev}. From the viewpoint of the effective supergravity,  the 
existence of different notions  of 
 'charge'  can be attributed to the
Chern-Simons terms  in the action \cite{K1,K3}.

The fact that the flux of    
$F\simeq dA$  should be quantized was an
early source of confusion. 
 One can argue for this  
by requiring that  the $\sigma$-model on  a 
 worldsheet $\Sigma$ with boundary be well defined 
 \cite{KS,Eli,K2,K4,K5}.\footnote{I thank Cumrun Vafa for discussions
of this point.}
Let  $\Sigma$ have the topology of a disk, and denote by the same symbol
its spacetime embedding. The boundary $\partial\Sigma$ is a loop
inside the worldvolume of some D-brane. Let  ${\cal D}$ be a disk inside
 this worldvolume such that $\partial\Sigma = \partial {\cal D}$ , and let
 ${\cal M}_3$  be a 3-manifold bounded by
 the two disks,  $ \partial{\cal M}_3=  \Sigma - {\cal D}$
(the minus sign refers to inverse orientation). This is illustrated in
 figure 1.
The  $\sigma$-model action can now  be written as follows:
\begin{equation}
S =  \int_\Sigma  {\rm tr}\; \hat G
+ \int_{\cal D} {\cal F} + \int_{{\cal M}_3}  H \ , 
\label{action}
\end{equation}
where $\hat G$ is the pull-back metric, 
 $H\simeq dB $ is the NS 3-form,
and ${\cal F}$ is  the gauge-invariant field on the D-brane. 
The measure is independent of the choice of  ${\cal M}_3$ provided
the $H$-flux through the three-sphere is quantized, $\int H = k+2$. 
Likewise, quantization of the $F$-flux threading any closed two-manifold
ensures that the measure is independent of  the choice of ${\cal D}$. 
Explicitly, given two disks ${\cal D}$ and ${\cal D}^\prime$, and a 
 `ball' ${\cal B}_3$ such that $\partial{\cal B}_3 = {\cal D}-{\cal D}^\prime$, 
we must have
\begin{equation}
\int_{{\cal D}-{\cal D}^\prime} {\cal F} -  \int_{{\cal B}_3}
 H\  =\  n\in Z\ .
\end{equation}
The ambiguity in the choice of ${\cal B}_3$ implies that 
we can only associate a 
 flux in 
$Z_{k+2}$  to any  closed two-manifold 
${\cal D}-{\cal D}^\prime$ on the D-brane.

\FIGURE{\begin{picture}(300,150)(0,0)
\put(30,0){\epsfig{file=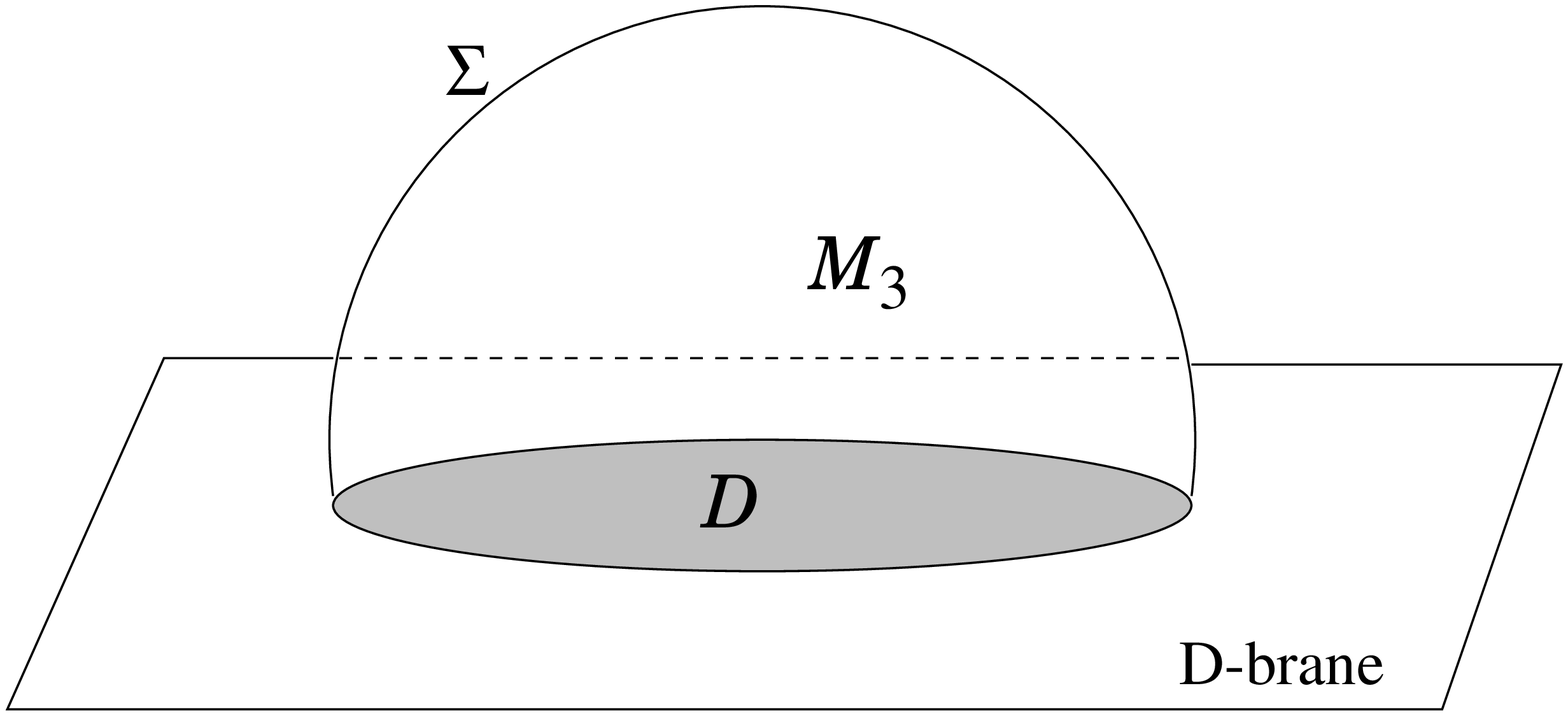,height=4cm}}
\end{picture}
\caption{ The spacetime image of the worldsheet $\Sigma$,  and a
disk ${\cal D}$ on the D-brane worldvolume, which together
 form the boundary of a 3-manifold
${\cal M}_3 $. These  enter in the $\sigma$-model
action for open strings, equation \eqref{action}. 
\label{orientiffolds}}}

 One can think of
the  integer $n$ as the number of D-particles, 
  which expand out in the
background $B$-field by a
 generalized `dielectric' effect \cite{Myers}.
The boundary RG flow,  describing the formation of the $n$-particle
bound state,  is the same  that describes  the screening of a
magnetic-impurity  spin  $s=(n-1)/2$  by $k$ species of conduction
electrons (the Kondo problem)  \cite{Kondo,ARS}. 
The fact that $n \simeq k+2-n$ can be understood
differently  in various string theory contexts.
Consider for instance 
 $Q_5 \equiv  k+2$ NS5-branes,  and a  D3-brane extending in
the transverse dimensions. The D-strings stretching between the
D3-  and NS5-branes have a chiral, fermionic  ground state, so that
pulling the D3-brane through the fivebranes  creates (destroys)
$k+2$  oriented D-strings \cite{HW,cr1,cr2,cr3}.  
In the  near-horizon geometry of the fivebranes, 
this process changes precisely $n$ to $k+2-n$ \cite{Pelc}.

A related argument, explained to me by Juan Maldacena, 
starts with 
a D3-brane wrapping the  SU(2) manifold.
 The DBI action includes  a term
\begin{equation}
\int B\wedge \;^*F = - \int H\wedge {\tilde A}\ , 
\label{baryo}
\end{equation}
with ${\tilde A}$ the (magnetic) gauge field dual to $A$.   
Equation \eqref{baryo} 
shows that the  $H$-flux  induces $k+2$ units of magnetic charge, 
that  can  be cancelled by $k+2$ D-strings 
ending on the D-brane. This is  analogous to the baryon
vertex in  AdS5 \cite{baryon}.
Now think of the  D-strings as Euclidean  D-particle trajectories,
which terminate  on a spherical  D-brane at fixed (Euclidean) time.
This  is  an instanton configuration, describing a process in
which  $k+2$ units
of D-particle charge disappear.\footnote{Lifted to M theory,
 such a process can  describe  the  elastic scattering of 
a Kaluza-Klein graviton transfering $Q_5$ units of momentum
to a bound state of fivebranes. I thank Ed Witten for a discussion of this point.} 
D-particle number is thus only conserved modulo $k+2$, 
as advertized.

\boldmath
\section{D-branes in SL(2,R)}
\unboldmath

In the SL(2,R) WZW model we lack, at present, an algebraic
construction of D-branes as conformal boundary states {\`a} la
Cardy (see, however, the recent work \cite{GKS} for some steps
in this direction). What has been
worked out are  the possible `gluing conditions' for 
worldsheet currents, and their
semiclassical interpretation 
based on the DBI action \cite{S1,BP}.
Let me  summarize very briefly the  results (the reader should
consult the references for details):

(a) The  SL(2,R) group elements  can be written 
\begin{equation}
g  = {1\over L}
\left(
\begin{array}{lll}
X^0+X^1 &\quad & X^2+X^3\cr
X^2-X^3 &\quad & X^0- X^1\cr
\end{array}
\right) ,
\label{param}
\end{equation}
with the $X^M$ parametrizing  a pseudosphere 
in flat space of  signature  $(-++-)$. The group manifold
 is Lorentzian AdS3 of radius $L$.
Identifying the currents by an inner automorphism, as in \eqref{gl}, 
 leads to two generic types of D-branes 
corresponding to the  elliptic or hyperbolic
   conjugacy classes of the group \cite{S1}. 
Their  geometry is two-dimensional de Sitter (dS2)
   or the two-dimensional hyperbolic plane (H2).
 There are in addition two  special cases: 
   pointlike D-instantons (the conjugacy class of the identity)
and the half  lightcones in AdS3.

(b) A third generic class of D-branes is obtained for the gluing
conditions \cite{BP}:
\begin{equation}
J^a T_a \; =\;  {\bar J^a}\; \omega_0^{\vphantom 1} T_a {\omega_0^{\vphantom 1}}^{-1}
\ \ \ \ {\rm with} \ \ \ \
\omega_0^{\vphantom 1} = \left(
\begin{array}{ll}
0 & 1 \cr
1 & 0 \cr
\end{array}
\right) .
\end{equation}
Here $T_a$ are the generators of the group, and $\omega_0^{\vphantom 1}$ is an
outer automorphism. By an  extension 
of the argument of the previous section,  one can show that these  D-brane
worldvolumes are `twined conjugacy classes' \cite{Fr}  for which  
${\rm tr}(\omega_0^{\vphantom 1} g)$ is fixed. Their  geometry is
two-dimensional anti-de-Sitter (AdS2). The various D-branes preserving a
SL(2,R) symmetry are collected in the table below. There are, in addition, 
stable non-symmetric branes that I wont discuss.

\vskip 0.2cm

\begin{table}[htp]
\begin{center}
\begin{tabular}{|c|c|}
\hline
&
\\
 Conjugacy class    &  D-brane 
\\
&
\\
\hline\hline
&
\\
 $-\infty < \tr (\omega_0 g) <\infty$   & AdS$_2$ 
\\
&
\\
\hline
&
\\
$\vert \tr g \vert < 2$   & $H_2$ 
\\
&
\\
\hline
&
\\
$\vert \tr g \vert > 2$ & dS$_2$  
\\
&
\\
\hline
&
\\
 $\vert \tr g \vert =  2$  & light cone  
\\
&
\\
\hline
&
\\
 $g =   {\bf 1}$  & point
\\
&
\\
\hline
\end{tabular}
\end{center}
\caption{
The different regular and twined  conjugacy classes of $SL(2,\bR)$,  and
the geometry of the corresponding D-brane worldvolumes.}
\end{table}

(c) The semiclassical analysis \cite{BP} shows that only the AdS2
branes are `physical'.
The dS2 geometries,  in particular,  correspond to 
motions of closed D-strings carrying a
supercritical electric field. This is in line
with 
other statements \cite{no1,no2}
about the impossibility to realize de Sitter
spaces in string theory. The AdS2 branes, on the other hand, 
are worldvolumes  of static (p,q)
strings stretching between antipodal points on the AdS boundary.
Two of those  are drawn schematically in figure 2. 
The larger the value of $q/p$,  the more the string  bends  towards
the AdS boundary.

\FIGURE{\begin{picture}(320,150)(0,0)
\put(30,0){\epsfig{file=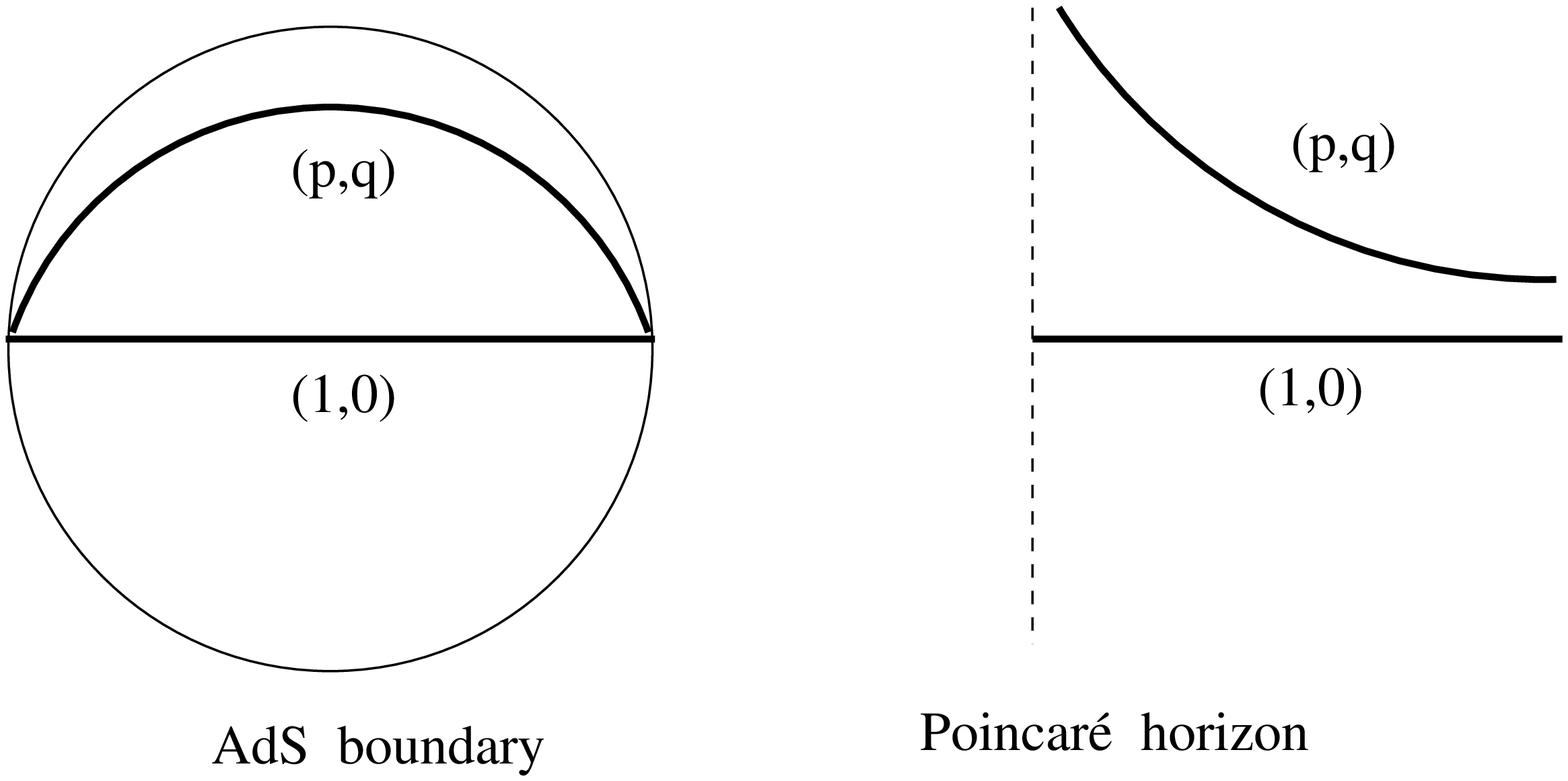,height=5cm}}
\end{picture}
\caption{The AdS2 D-branes in cylindrical (left) and Poincar{\'e} coordinates (right).
The straight brane is the worldvolume of a pure   D-string. Binding $q$ fundamental
strings to it, makes it bend towards the boundary of the ambient AdS3 spacetime.  
\label{ads}}}

\vskip 0.3cm

 (d) All of these branes
have a supersymmetric embedding \cite{S1,BP}
  in the
 AdS3$\times$S3
$\times$(T4 or K3) 
background arising in the near horizon region
of a NS5/F1 black string \cite{ms}.
For the S3 component of the space, one must use the SU(2)
branes of the previous section. The
 AdS2$\times$S2 branes, in particular, describe the
junction of a (p,q) string with the NS5/F1 string of the
background \cite{BP}. Adding momentum along  this latter string amounts
to modding out AdS3 by a discrete isometry, exactly as for
the BTZ black hole. This 
breaks all  the supersymmetries of the D-brane.

   The worldvolume theory of the AdS2$\times$S2 branes is an interesting
deformation of N=4 super Yang-Mills. The theory has  N=2 supersymmetries,
and it approaches in appropriate limits 
the  non-commutative
theory on the `fuzzy' sphere \cite{ARS}, 
and a curved 
version of the NCOS theory \cite{NCOS,NCO}. 
It would be interesting to study S-duality in this context. 
Steps towards deriving the full open-string
spectrum on the AdS2 branes 
have been taken recently in references \cite{PR,Hi}.
In the case $q=0$, in particular, the spectrum  is the  `holomorphic
half' of the closed-string spectrum 
proposed for the SL(2,R) WZW model
in reference \cite{mo}.

  Another very interesting question concerns  the  interpetation
of the AdS  D-branes in the dual spacetime CFT. Their holographic
`images'   turn  out to be 
conformal defects separating  different CFTs on either side
\cite{KR2,prep}. One can also extend  the 
geometric  considerations of this talk
so as to incorporate  orientifolds \cite{prep1},
recovering in particular the algebraic results of \cite{or1,or2}.
I will not discuss these issues here any further --
I will  zoom instead, in the concluding section,
on the  intriguing  interplay of
effective versus  induced geometry.

\vfil\eject

\boldmath
\section{Brane worlds and `radius locking'}
\unboldmath

   The AdS2$\times$S2  branes of the previous
 section have higher-dimensional
analogs in other near-horizon geometries. A  particularly
interesting example  are the AdS4$\times$S2 branes in the
AdS5$\times$S5 geometry with five-form Ramond Ramond background. 
These have been analyzed recently by Karch and Randall \cite{KR2,KR1},
as a step towards the realization of warped `brane world' 
compactifications in string theory. The question of gravity localization
requires to go beyond the probe approximation, and to study the
back reaction of the branes on the ambient geometry. 

 Another potential obstruction to the Randall-Sundrum
scenario, of purely string-theoretic  origin, has been pointed out in
reference \cite{BP}. It
 has to  do with the fact that, in the
presence of non-trivial fluxes, the induced and effective metrics on the brane
can be drastically different. Let me now explain this in some more detail. 

 The radius of the AdS2 branes of the previous
section, measured in  the induced (closed-string) metric is
\begin{equation}
{\hat l}_{\rm AdS2}\; = \; L { T_{(p,q)}\over T_{(p,0)} }\; \ge \; L \  ,  
\end{equation}
where $T_{(p,q)}$ is the tension of a (p,q) string,  which is 
always greater or equal
than  the tension of $p$ pure D-strings. By taking $q\to\infty$
one can make ${\hat l}_{\rm AdS2}$ arbitrarily large, 
so that the brane is much more flat than the ambient geometry. 
The S2 part of the brane, on the other hand, cannot be bigger than the
equator two-sphere. A straightforward calculation gives 
\begin{equation}
{\hat l}_{\rm S2}\; =\; L\;\sin\left({\pi q\over k+2}\right)\; \le\;  L , 
\end{equation}
so that the spherical brane can be arbitrarily more curved than the
background geometry.
 This situation seems,  at
first sight, paradoxical because  unbroken supersymmetry requires
the AdS2 and S2 radii to be equal. The point, however, is that the
Yang-Mills multiplet couples to 
 an effective open-string metric,  which is related to
the closed-string metric through the well-known  formula \cite{Ab,SW}
\begin{equation}
G_{\alpha\beta} =
 \hat g_{\alpha\beta} - {\cal F}_{\alpha\gamma}\; \hat
g^{\gamma\delta}\; 
{\cal F}_{\delta\beta}\ .
\end{equation}
Measured in the open-string metric the two radii are,  indeed,  equal to each other
and to the background radius (which is the same for AdS3 and S3), 
\begin{equation}
{L}_{\rm S2}\; =\; {L}_{\rm AdS2}\; = L \ ,
\label{lok} 
\end{equation}
for all values of p and q. This `locking' of the effective radii 
to the ambient values has been
observed also in a related context in \cite{MMS}. 
One could have argued for it from the fact that the
open- and closed-string spectra are related.
Note that the causal structure is  determined by the
 lightcones of the closed-string metric, 
in accordance with the
general arguments of \cite{GH}. In other words nothing travels faster
than gravitons.

Relation \eqref{lok} shows that it is impossible to  fine-tune the
 {\it effective} geometry of the D-brane to be  `flatter'
 than the bulk geometry.
 It is unclear whether this  phenomenon  persists 
in higher dimensions and for
 Ramond-Ramond fluxes. The back-reaction of the branes
on the ambient metric
could also play an important role. 
If the above conclusion were, however,  to persist,  it
 would be an  obstruction to
the construction of  realistic `warped brane worlds'
in string theory.  

\vskip 0.3cm

{\bf Aknowledgements}: I am grateful to  the organizers for the invitation
to speak. I  thank Mike Douglas, Marios Petropoulos and Christoph
Schweigert for very pleasant collaborations on the contents of this
talk. Finally, I aknowledge the
support of the European Networks ``Superstring theory'' (HPRN-CT-2000-00122)
and ``the quantum structure of spacetime'' (HPRN-CT-2000-00131).


\end{document}